\documentclass[12pt]{article}
\usepackage{graphicx}
\usepackage{multirow}
\usepackage{amsmath}

\newcommand\pubnumber{ }
\newcommand\pubdate{\today}

\def\napoli{Department of Physics and Astronomy\\
University of Hawai`i at M\={a}noa\\
2505 Correa Road\\
Honolulu, HI, USA 96822.}
\def\support{\footnote{On behalf of the Belle~II Collaboration.}}

\def\Title#1{\begin{center} {\Large #1 } \end{center}}
\def\Author#1{\begin{center}{ \sc #1} \end{center}}
\def\Address#1{\begin{center}{ \it #1} \end{center}}

\newcommand\pubblock{\rightline{\begin{tabular}{l} \pubnumber\\
         \pubdate  \end{tabular}}}
\newenvironment{Abstract}{\begin{quotation}  }{\end{quotation}}
\newenvironment{Presented}{\begin{quotation} \begin{center} 
             PRESENTED AT\end{center}\bigskip 
      \begin{center}\begin{large}}{\end{large}\end{center} \end{quotation}}

\def\beq{\begin{equation}}
\def\eeq#1{\label{#1}\end{equation}}
\def\eeqn{\end{equation}}

\def\beqa{\begin{eqnarray}}
\def\eeqa#1{\label{#1}\end{eqnarray}}
\def\eeqan{\end{eqnarray}}

\let\bar=\overbar

\def\Dslash{\not{\hbox{\kern-4pt $D$}}}
\def\dslash{\not{\hbox{\kern-2pt $\del$}}}

\def\msb{{\bar{\ssstyle M \kern -1pt S}}}

\begin{document}
\begin{titlepage}
\pubblock

\vfill
\Title{New Physics Prospects in Mixing and CP Violation at Belle~II}
\vfill
\Author{ Kurtis Nishimura\support}
\Address{\napoli}
\vfill
\begin{Abstract}
The Belle~II experiment at the SuperKEKB electron-positron collider will 
provide a large sample of charm mesons in addition to its primary goal of
$B$ meson production.  The large data sample and wide variety of accessible 
$D$ meson decay modes in a clean experimental environment provide sensitivity
to new physics via $D$ meson mixing and CP violation measurements.  This 
contribution briefly describes selected components of the upgrade from Belle to 
Belle~II and KEKB to SuperKEKB, and some prospects for new physics searches in 
the areas of charm mixing and CP violation. 
\end{Abstract}
\vfill
\begin{Presented}
The 5$^{\rm{th}}$ International Workshop on Charm Physics\\
Honolulu, Hawaii, USA, May 14-17, 2012
\end{Presented}
\vfill
\end{titlepage}
\def\thefootnote{\fnsymbol{footnote}}
\setcounter{footnote}{0}

\section{Introduction}

In addition to their primary mission of exploring CP violation and
other properties of the $B$ meson system,
the $B$ factory experiments Belle and BaBar have produced a wealth of
measurements in the charm sector.  This is in part due to the large 
cross section for charm production at these facilities.  For example, 
at the nominal running energy of $10.58~\mathrm{GeV}$, the cross section 
for $e^+e^- \to c\bar{c}$ is actually larger than that of the nominal
process, $e^+e^- \to \Upsilon(4S) \to B\bar{B}$.  Cross sections for
these two processes are $1.3~\mathrm{nb}$ and $1.1~\mathrm{nb}$, 
respectively.  Thus, over a decade of operation, Belle alone has 
collected $\approx10^9$ charm events in over $1~\mathrm{ab}^{-1}$ 
of integrated luminosity.  This data set, combined with those taken
by CLEOc, BaBar, LHCb, and others, are allowing increasingly precise
probes for new physics in charm mixing and CP violation.

However, the nature of $D$ meson mixing and CP violation are such
that the effects are expected (and thus far observed) to be quite small.
To make significant headway in future studies on these topics will
require a substantial increase in statistics.  Belle~II and SuperKEKB
are ongoing upgrades to the Belle detector and KEKB asymmetric $e^+e^-$ 
collider, respectively, and will provide the necessary experimental 
precision to push these searches for new physics to the next level.

This contribution briefly introduces some $D$ mixing and CP violation formalism, and
describes measurements of these parameters that are particularly well suited
for study at $e^+e^-$ experiments.  The current experimental status of selected 
measurements in these sectors is then described, and prospects for improved 
measurements with Belle~II at SuperKEKB are presented.

\section{$D$ meson mixing and CP violation}

As in other neutral meson systems, mixing occurs when the flavor and 
mass eigenstates are not identical.  For the neutral $D$ system, this
is usually written as: $|D_{1,2}> = p |D^0> \pm q | \bar{D^0}>$, where 
$D_{1,2}$ are the mass eigenstates and $D^0$, $\bar{D^0}$ are the flavor
eigenstates.  The parameters $p$ and $q$ must satisfy $|p|^2 + |q|^2 = 1$, 
and, if CP is conserved, $p = q$.  
In the $D$ system, the mixing parameters are usually defined
as $x \equiv \Delta m / \Gamma$ and $y \equiv \Delta \Gamma / \Gamma$,
where $\Delta m$ and $\Delta \Gamma$ are the mass and decay width 
differences of the mass eigenstates, and $\Gamma$ is the average decay width.

In the Standard Model, neutral $D$ meson mixing is both CKM and GIM 
suppressed.  For example, the parameter $x$ is proportional to 
$(m_s^2 - m_d^2)/m_W^2$.  Typical standard model expectations for both
mixing parameters $|x|$ and $|y|$ are $\mathcal{O}(10^{-3})$ to 
$\mathcal{O}(10^{-2})$.  The small expected values of the mixing 
parameters within the Standard Model make them an excellent place to 
search for new physics, as observation of large mixing would require 
non-Standard Model explanations.

CP violation is also expected to be quite small within the Standard Model, 
where the CKM element responsible is $V_{cs} \sim \eta A^2 \lambda^4$, which
is roughly $\mathcal{O}(10^{-3})$.  This makes it another appealing place
to search for new physics, which could manifest itself through relatively
large values of CP asymmetry.  The asymmetry can be separated into three
pieces:
\[A_{CP}(D \to f) = \frac{\Gamma(D \to f) - \Gamma(\bar{D} \to \bar{f})}{\Gamma(D \to f) + \Gamma(\bar{D} \to \bar{f})} = a_f^m + a_f^d + a_f^i \mathrm{.}\]
The first contribution to the asymmetry, $a_f^m$, is that of mixing, and 
is nonzero when the mixing parameter $p \neq q$, as described previously.
The second contribution, $a_f^d$, is a direct component, where the decay
amplitudes of $D$ and $\bar{D}$ to final states $f$ and $\bar{f}$ are 
unequal.  The last contribution, $a_f^i$, comes from interference between
the mixing induced and direct CP violation.  This contribution is usually
represented as an angle $\phi$, defined as:
\[\phi = \arg\frac{q\mathcal{A}(\bar{D} \to \bar{f})}{p\mathcal{A}(D \to f)} \mathrm{.}\]

Searches for CP violation can thus be conducted in both time-dependent 
mixing measurements as well as time-integrated measurements of direct
CP asymmetries.  Both types of measurements are well-suited to 
$e^+e^-$ environments.

\section{$D$ meson production and measurement}

At the $B$ factories, charm mesons are copiously produced as part of the 
continuum process $e^+e^- \to c\bar{c}$.  A typical experimental
technique to identify neutral $D$ mesons is to use the decay process
$D^{*\pm} \to \pi_s^\pm D^0$, where $\pi_s^\pm$ is a characteristic 
``slow" pion.  By measuring the slow pion charge, the flavor of the
neutral $D$ meson is identified.   The decay length of the $D^0$ can
be measured with a vertex fit to the $D^0$ decay products, and the 
decay length can thus be inferred as the distance from the decay vertex
to the interaction region.  Decay time distributions and asymmetries
can thus be measured for a wide variety of states.

While it is true that similar measurements can be made at hadron collider
experiments, such as those at the LHC, the $e^+e^-$ environment is 
significantly cleaner.  This is more suitable for final states that include 
neutral particles (e.g., $\pi^0$ or $\eta$).  
Furthermore, since the full center-of-mass energy of each event is also known,
remaining particles not associated with the signal $D^* \to D \to f$ decay 
can be reconstructed as part of a tag meson, $D_{\mathrm{tag}}$.  In this way,
it is possible to identify signal decay chains with missing energy. 

Since the $B$ factories usually operate at or above the $B\bar{B}$ threshold,
there is a potential background from $D^*$ mesons produced from $B$ meson 
decays.  However, these backgrounds are kinematically well separated from 
continuum $D^*$ mesons, and do not significantly impact the analysis once
$D^*$ momentum criteria are imposed.

\section{Belle~II at SuperKEKB}

The KEKB $B$ factory began physics operation in 1999.  In over a decade 
of running, KEKB delivered a peak luminosity of 
$2.1~\times~10^{34}~\mathrm{cm}^{-2}\mathrm{s}^{-1}$ and a 
total integrated luminosity of over $1~\mathrm{ab}^{-1}$, allowing the 
Belle detector to accumulate a large experimental data set, primarily
at the energy of the $\Upsilon(4S)$ resonance.  

In 2010, KEKB ceased operation to begin upgrading to SuperKEKB, which is 
expected to provide the upgraded Belle~II detector with more than 
$50~\mathrm{ab}^{-1}$ by 2022.  Details on both the SuperKEKB and 
Belle~II upgrades can be found elsewhere \cite{TDR}, but two specific 
system upgrades are briefly described, as they are particularly relevant to 
charm studies.

Precision tracking at Belle was provided by a silicon vertex detector (SVD),
consisting of four layers of double sided silicon strip detectors 
\cite{SVD2}. At Belle~II, this system will be replaced with two layers 
of DEPFET pixel detector, surrounded by a four layer SVD.  The inner radius 
of the precision tracker thus decreases from 20~mm to 13~mm, while the outer 
radius increases from 8.8~cm to 13.5~cm.  The improved resolution of the pixels 
and the overall increase in coverage and acceptance correspond to an expected
$\sim25\%$ improvement in vertex resolution and $\sim30\%$ gain in efficiency 
to reconstruct the displaced vertex of $K_S^0 \to \pi^+\pi^-$.

Belle~II also benefits from significant upgrades in particle identification (PID) 
capabilities.  Belle utilized threshold aerogel Cherenkov detectors
and time-of-flight counters as dedicated PID devices.  Belle~II replaces these 
systems with two ring imaging Cherenkov devices: a time-of-propagation detector 
in the barrel, and an aerogel RICH in the forward endcap.  
The significant improvements in $K\pi$ separation reduce kinematic reflections.
For example, the mode $D^0 \to \pi^-\pi^+\pi^0$ suffers some background
from $D^0 \to K^-\pi^+\pi^0$ when the kaon is misidentified as a pion.  The 
improved particle idenfication at Belle~II should help to reduce uncertainties
due to these effects. 

\section{Mixing and Mixing-Induced CP Violation}

Measurements of $D^0\bar{D^0}$ mixing have been performed in a variety of
final states, with evidence for mixing observed by Belle, BaBar, and CDF.
While the mixing parameters of ultimate interest are the $x$ and $y$ defined 
previously, each final state allows access to these parameters in 
different combinations.  Some modes, such as 
$D^0 \to K_S^0 \pi^+ \pi^-$, allow direct access to $x$, $y$, $|q/p|$, 
and $\phi$.  In many cases, the observables are combinations of these
underlying parameters.
For example, the modes $D^0 \to K^+K^-$ and $D^0 \to \pi^+\pi^-$ 
allow measurement of $y_{CP}$ and $A_\Gamma$, which are defined as:
\begin{align*}
y_{\mathrm{CP}}&=y\cos(\phi) - \frac{1}{2}A_M x \sin\phi \\
A_\Gamma &= \frac{1}{2} A_M y \cos\phi - x \sin\phi \mathrm{,}
\end{align*}
with $A_M$ defined as
\[A_M = \frac{|q/p|^2 - |p/q|^2}{|q/p|^2+|p/q|^2} \mathrm{.}\]

\begin{figure}
\centering
\begin{tabular}{cc}
\includegraphics[width=0.46\linewidth]{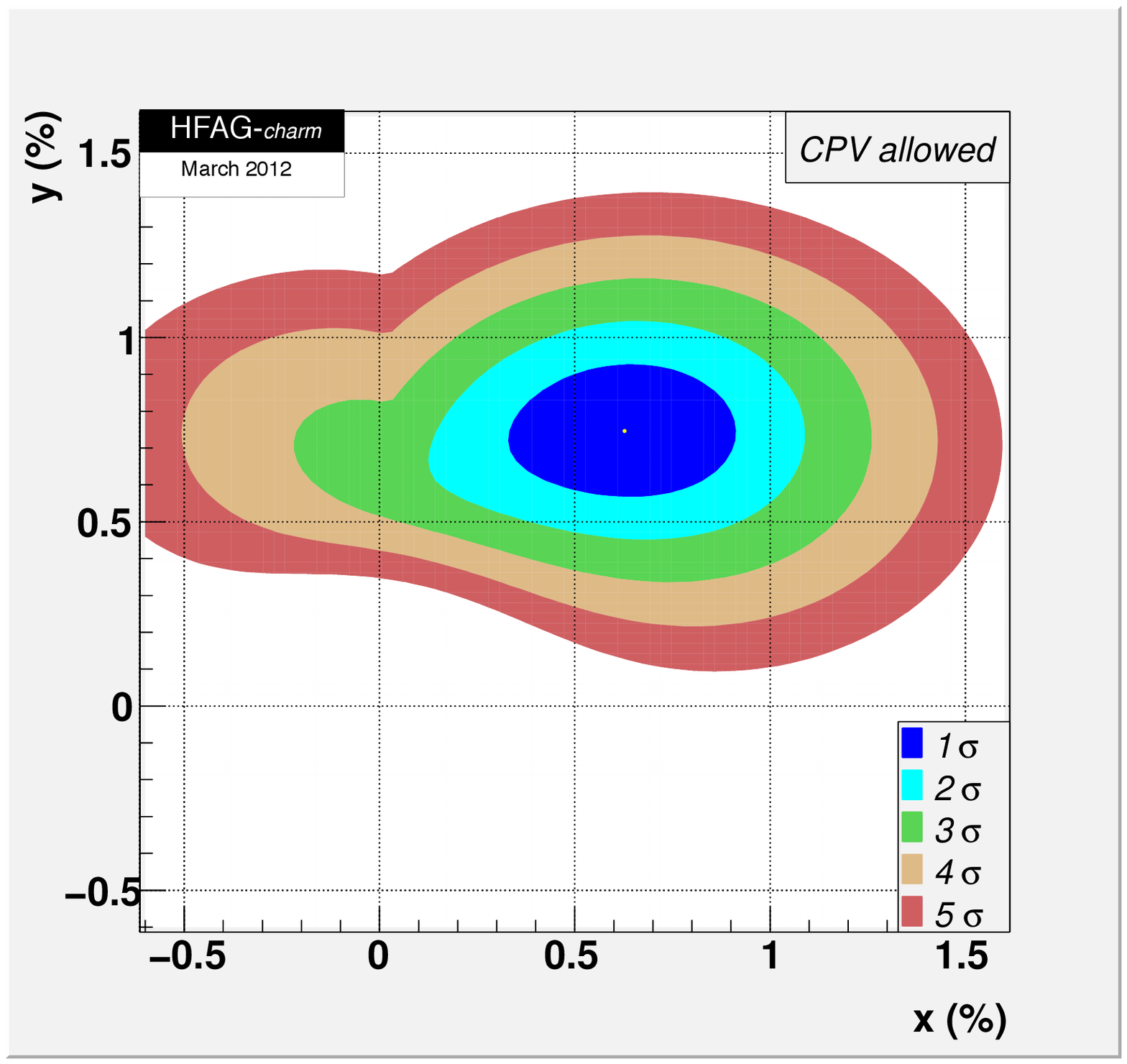} & 
\includegraphics[width=0.46\linewidth]{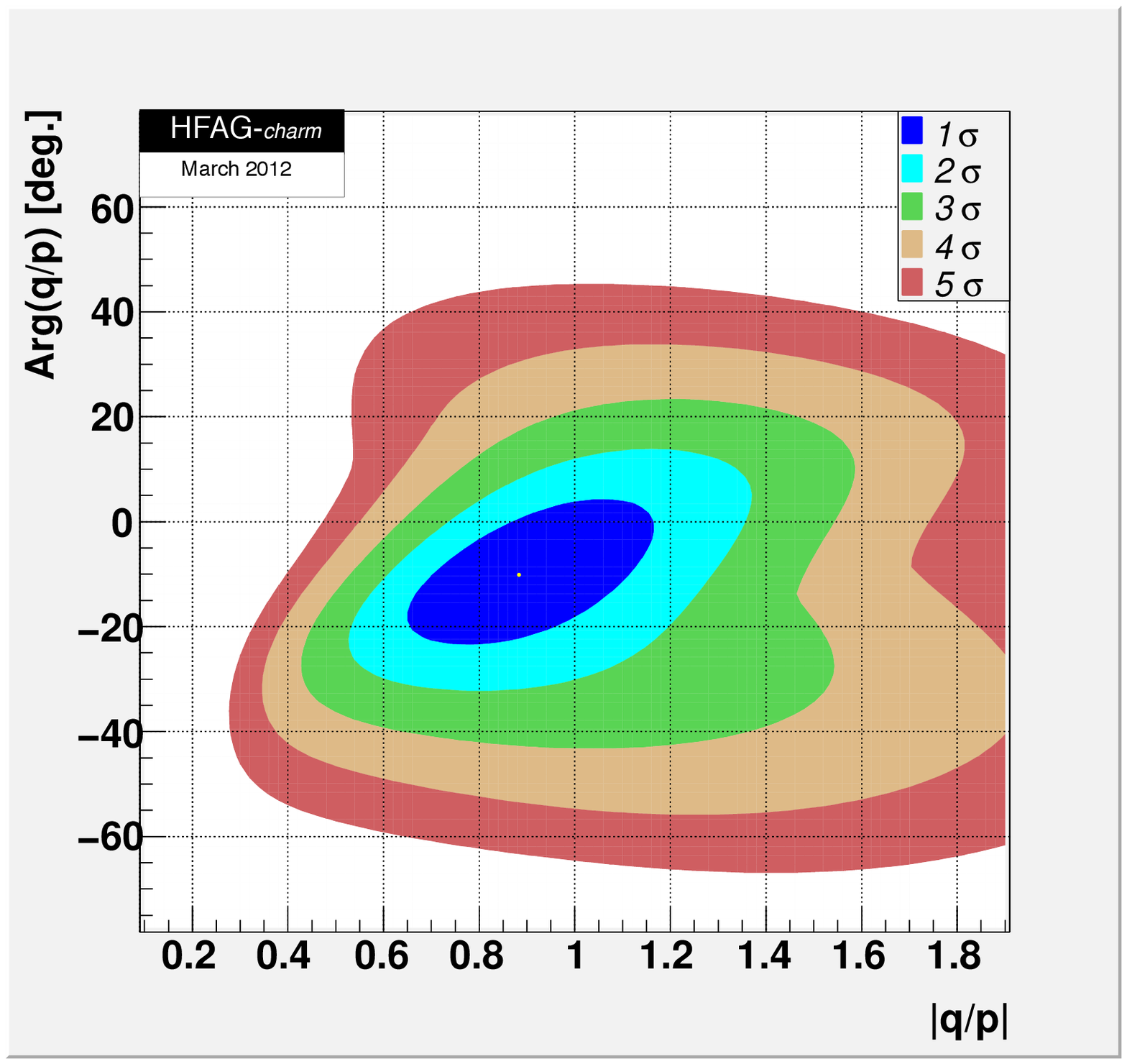}\\
\end{tabular}
\caption{(Left) Results of the current global fit to charm mixing parameters, 
allowing for CP violation.  The origin corresponds to the no-mixing point. 
(Right) Global fits to the current CP violation parameters.  The no-CP
violation point corresponds to the point (1,0).  Both plots are from 
Reference~\cite{HFAG}.}
\label{fig-HFAG-fit}
\end{figure}

Thus, in order to fully utilize the existing experimental data, information
from all modes is combined into a global fit for $x$, $y$, $\arg{(q/p)}$, 
and $|q/p|$ \cite{HFAG}.  The current global fit to these values is shown
in Figure \ref{fig-HFAG-fit}.  The fitted parameters and their uncertaintes are:
\begin{align*}
x     &= (0.63  \pm 0.19)\% \\
y     &= (0.75  \pm 0.12)\% \\
|q/p| &= (0.88  \pm 0.17) \\
\phi  &= (-10.3 \pm 9.2)^\circ \mathrm{.}
\end{align*}
While the no-mixing hypothesis is excluded at high statistical significance
$(>10\sigma)$, the observed levels of mixing are compatible with Standard Model
expectations.  The global fit remains compatible with the 
hypothesis of no CP violation.

The strength of the Belle~II charm physics program lies in its ability to 
access many final states in the same experiment.  This ensemble of data
allows for precision determinations of the mixing and CP violating 
parameters.
To see the expected effect of Belle~II measurements, a similar global fit 
is performed using projected sensitivities at $50~\mathrm{ab}^{-1}$ of 
integrated luminosity for the following modes: $K^+K^-$, $\pi^+\pi^-$, 
$K^+\pi^-$, and $K_S^0\pi^+\pi^-$.  The expected uncertainties for these 
measurements are shown in Table~\ref{tab-mixing-BelleII}.
In estimating these sensitivities, anticipated improvements from statistics 
are included, as well as improvements to systematic uncertainties that are 
expected to decrease with higher statistics.  

\begin{table}
\begin{center}
\begin{tabular}{|c|c|cc|}  
\hline
\multirow{2}{*}{$D^0$ Decay Mode} & \multirow{2}{*}{Observable} & \multicolumn{2}{|c|}{Uncertainty (\%)} \\
  &  & Current ($\sim1.5~\mathrm{ab}^{-1}$) & Belle~II ($50~\mathrm{ab}^{-1}$) \\
\hline
\multirow{4}{*}{$K_S^0\pi^+\pi^-$}   & $x$        & 0.211    & 0.10 \\
                                     & $y$        & 0.186    & 0.08 \\
                                     & $|q/p|$    & 32       & 9    \\
                                     & $\phi$     & 0.32 rad & 0.07 rad\\
\hline
\multirow{3}{*}{$\pi^+\pi^-/K^+K^-$} & $y_{CP}$   & 0.217    & 0.05\\
                                     & $A_\Gamma$ & 0.248    & 0.03\\
                                     & $A_{CP}$   & 0.240    & 0.07\\
\hline
\multirow{4}{*}{$K^+\pi^-$}          & $x'^2$     & 0.0195   & 0.009\\
                                     & $y'$       & 0.321    & 0.16\\
                                     & $A_{D}$    & 3.5      & 1.7\\
                                     & $R_{D}$    & 0.013    & 0.0015\\
\hline
\end{tabular}
\caption{Projected Belle~II sensitivities, relative to existing
world averages, for mixing and mixing-induced CP violation parameters in 
selected $D^0$ decay modes.}
\label{tab-mixing-BelleII}
\end{center}
\end{table}

\begin{figure}
\centering
\begin{tabular}{cc}
\includegraphics[width=0.46\linewidth]{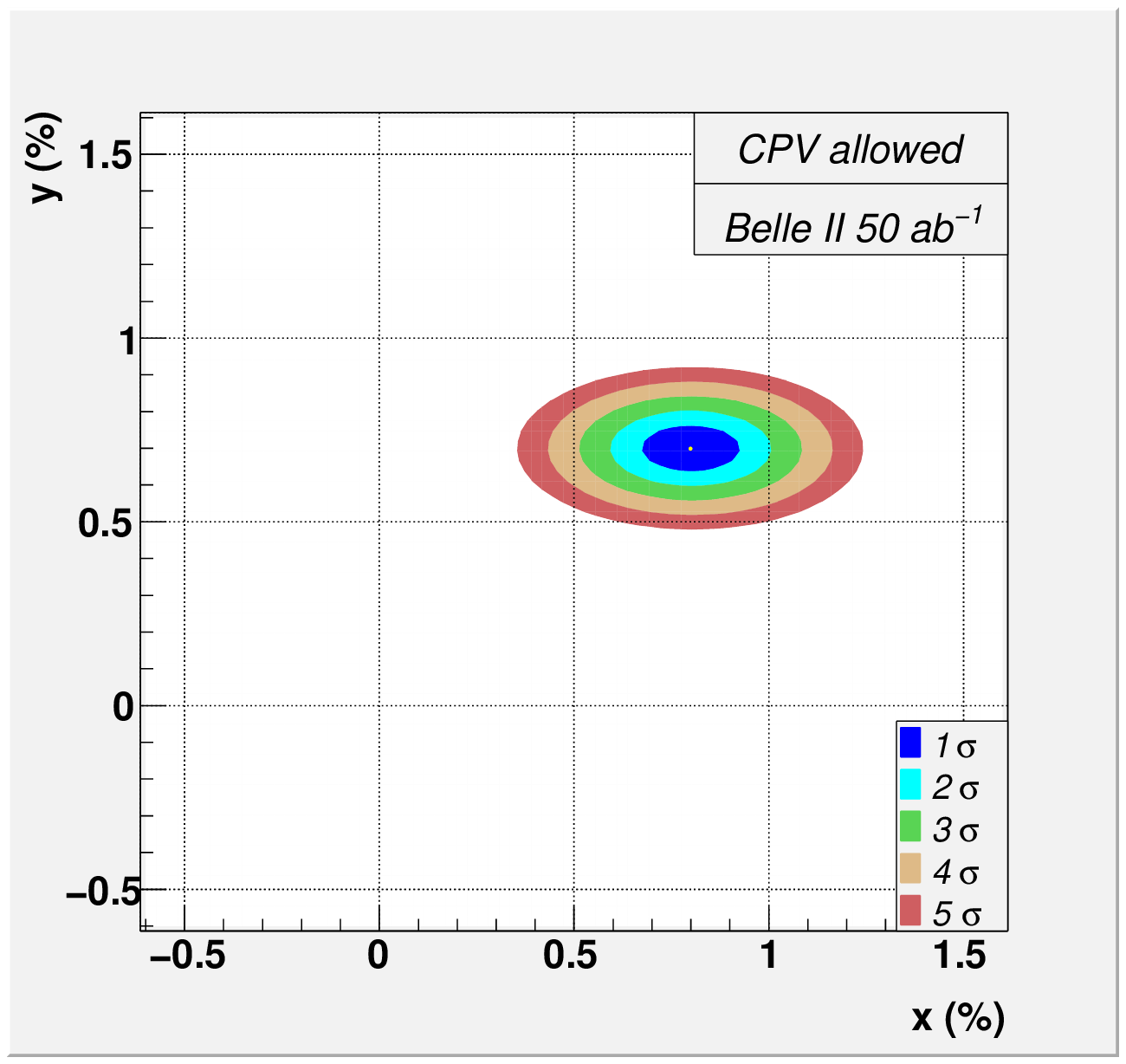} & 
\includegraphics[width=0.46\linewidth]{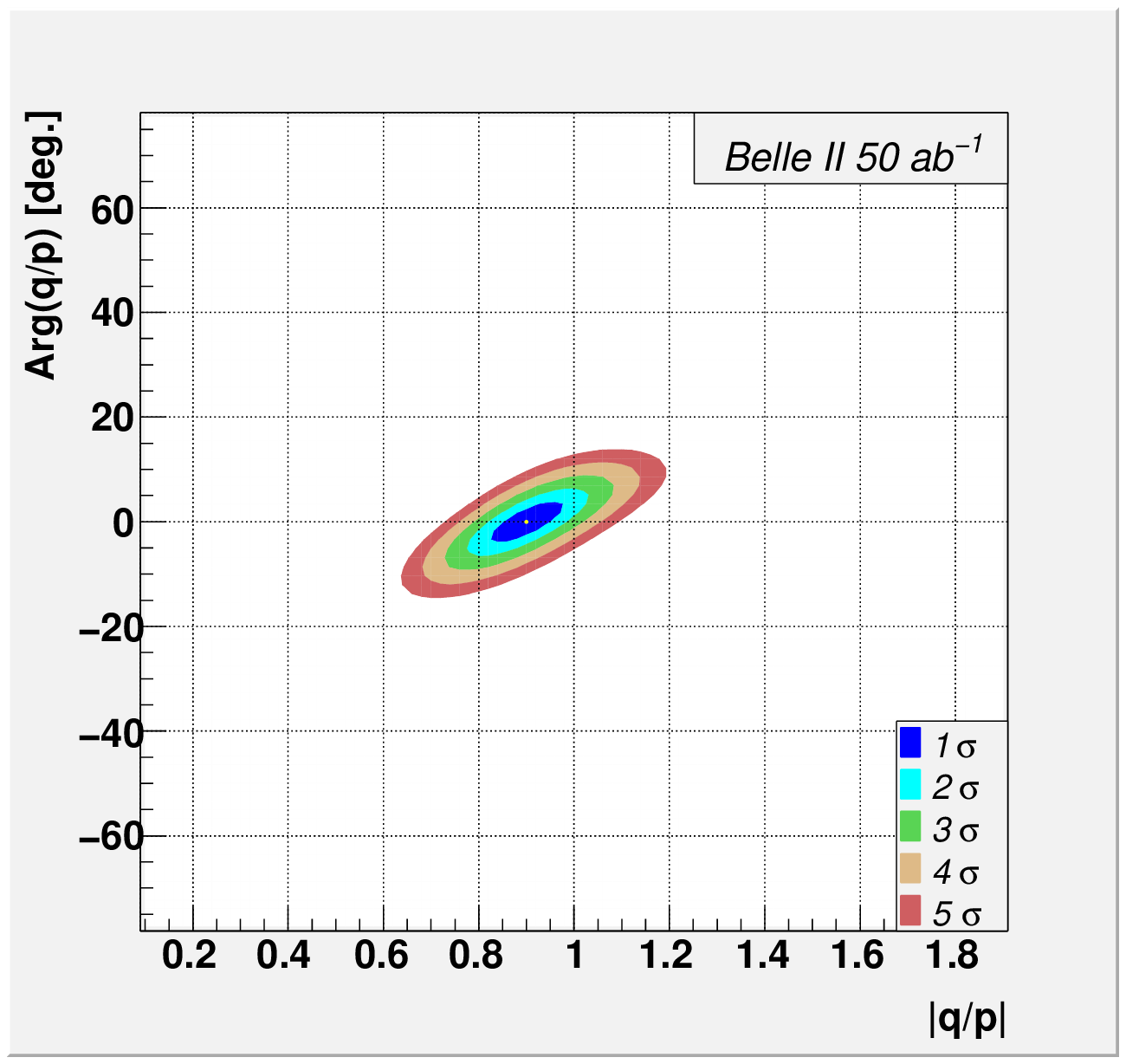}\\
\end{tabular}
\caption{Equivalent fits to Figure~\ref{fig-HFAG-fit}, but using only
Belle~II projected uncertainties for the $D$ decay modes included 
in Table~\ref{tab-mixing-BelleII}.}
\label{fig-HFAG-like-fit-BelleII}
\end{figure}

Projected results from these modes are combined into a global fit, 
with results shown in Figure~\ref{fig-HFAG-like-fit-BelleII} and 
Table~\ref{tab-HFAG-like-fit-BelleII}.  These and other Belle~II 
sensitivity projections can be found in Ref.~\cite{PhysSuperBFactory}.
Note that these results do not take into account all possible
decay modes, or include contributions from other experiments.  Even with 
these conservative sensitivities, the improvements are dramatic.
Sensitivity to both mixing and CP violation parameters is expected
to improve considerably.

\begin{table}
\begin{center}
\begin{tabular}{|c|c|c|}  
\hline
Parameter & Result of current global fit & Expected Belle~II precision \\
\hline
$x$     & $(0.63^{+0.19}_{-0.20})\%$    & $\pm0.08\%$ \\
$y$     & $(0.75\pm0.12)\%$             & $\pm0.04\%$ \\
$|q/p|$ & $0.88^{+0.18}_{-0.16}$        & $\pm0.05$ \\
$\phi $ & $(-10.1^{+9.5}_{-8.9})^\circ$ & $\pm2.6^\circ$ \\
\hline
\end{tabular}
\caption{Projected sensitivities to selected neutral $D$ meson mixing and CP
violation parameters, relative to results of the existing global fit.}
\label{tab-HFAG-like-fit-BelleII}
\end{center}
\end{table}

These types of measurements of mixing and CP violation parameters provide
powerful constraints on many scenarios involving new physics.  For example,
Reference~\cite{PRD76} utilizes the wide array of available $D$ mixing 
measurements to set constraints on models with a fourth generation of fermions,
extra gauge bosons, extra dimensions, and many others.  Thus, even if a 
"smoking gun" signal of large mixing-induced CP violation is not observed 
at Belle~II, the improvements to the ensemble of all $D$ mixing measurements 
will help to squeeze the parameter space of an extremely varied set of new 
physics models.  

\section{Time-integrated searches for CP violation}

In addition to the mixing-induced CP violation searches previously described,
it is also possible to search for direct CP violation using time-integrated
measurements.  The observable for direct CP violation searches is an
asymmetry in the decay widths of a decay mode and its charge conjugate:
\[A_{CP} = \frac{\Gamma(D \to f) - \Gamma(\bar{D} \to \bar{f})}{\Gamma(D \to f) + \Gamma(\bar{D} \to \bar{f})} \mathrm{.}\]
As with mixing and indirect CP violation, the expected level of 
$A_{CP}$ in the Standard Model is small, so observation of
large direct CP asymmetry in $D$ decays could be indicative of new physics.

Measurement of $A_{CP}$ can be contaminated by environmental effects.
Reconstruction asymmetries between positive and negative kaons or pions 
must be estimated and removed from the raw measurement.  Furthermore, the 
underlying production process may have an intrinsinc asymmetry.  At Belle
and Belle~II, for example, there is a forward-backward asymmetry from 
interference between $\gamma$ and $Z$ production channels.  This
contribution must be corrected out to determine the contribution from $D$ 
decays.

At the LHC, the underlying $pp$ initial state is not CP symmetric, making
absolute measurements of direct CP asymmetries prone to systematic
uncertainties.  However, these large uncertainties can cancel out if 
the difference in asymmetry between two modes is measured. 
LHCb has reported a difference in direct CP asymmetry between the final 
states $K^+K^-$ and $\pi^+\pi^-$ \cite{LHCb-Acp}.  Their result is 
$\Delta A_{CP} = [-0.82 \pm 0.21 \mathrm{(stat.)} \pm 0.11 \mathrm{(syst.)}]\%$,
with a significance of $3.5\sigma$.  This constitutes the first evidence
for direct CP violation in the $D$ meson system.  CDF has also
confirmed this result with their own measurement, 
$\Delta A_{CP} = [-0.62 \pm 0.21 \mathrm{(stat.)} \pm 0.10 \mathrm{(syst.)}]\%$, with a
significance of $2.7\sigma$ \cite{CDF-Acp}.

It is not yet clear if the LHCb result is consistent with the Standard
Model.  For example, some calculations indicate that Standard Model penguin
amplitude contributions to these processes may account for the result, 
and that the key to unraveling this contribution is to measure $A_{CP}$ 
in a variety of modes \cite{BGR}.  Given the measured value of $\Delta A_{CP}$
between the $K^+K^-$ and $\pi^+\pi^-$ modes, the authors of Ref.~\cite{BGR} 
predict non-zero CP asymmetries in the following decays:
$D^+ \to K^+\bar{K^0}$, $D^0 \to \pi^0\pi^0$, $D_s^+ \to \pi^+K^0$,
and $D_s^+ \to \pi^0K^+$.  They also predict null asymmetries for 
$D^+ \to \pi^+\pi^0$ and $D^0 \to K^0\bar{K^0}$.  It is again clear that to
make significant headway in our understanding of possible new physics in
this sector, measurements of CP asymmetries for as many modes as possible 
are required.

\begin{table}
\begin{center}
\begin{tabular}{|l|c|cc|}  
\hline
\multirow{2}{*}{Decay Mode} & \multirow{2}{*}{$\mathcal{L}[\mathrm{fb}^{-1}$} & \multicolumn{2}{|c|}{$A_{CP}$ (\%)} \\
  &  & Belle & Belle~II ($50~\mathrm{ab}^{-1}$) \\
\hline
$D^0 \to K_S^0\pi^0$         & 791 & $-0.28 \pm 0.19 \pm 0.10$ & $\pm 0.05$\\
$D^0 \to K_S^0\eta$          & 791 & $+0.54 \pm 0.51 \pm 0.16$ & $\pm 0.10$\\
$D^0 \to K_S^0\eta^\prime$   & 791 & $+0.98 \pm 0.67 \pm 0.14$ & $\pm 0.10$\\
$D^0 \to \pi^+\pi^-$         & 540 & $+0.43 \pm 0.52 \pm 0.12$ & $\pm 0.07$\\
$D^0 \to K^+K^-$             & 540 & $-0.43 \pm 0.30 \pm 0.11$ & $\pm 0.05$\\
$D^0 \to \pi^+\pi^-\pi^0$    & 532 & $+0.43 \pm 1.30$          & \\
$D^0 \to K^+\pi^-\pi^0$      & 281 & $-0.6  \pm 5.3$           & \\
$D^0 \to K^+\pi^-\pi^+\pi^-$ & 281 & $-1.8  \pm 4.4$           & \\
\hline
$D^+ \to \phi\pi^+$          & 955 & $+0.51 \pm 0.28 \pm 0.05$ & $\pm 0.05$\\
$D^+ \to \eta\pi^+$          & 791 & $+1.74 \pm 1.13 \pm 0.19$ & $\pm 0.20$\\
$D^+ \to \eta^\prime\pi^+$   & 791 & $-0.12 \pm 1.12 \pm 0.17$ & $\pm 0.20$\\
$D^+ \to K_S^0\pi^+$         & 673 & $-0.71 \pm 0.19 \pm 0.20$ & $\pm 0.05$\\
$D^+ \to K_S^0K^+$           & 673 & $-0.16 \pm 0.58 \pm 0.25$ & $\pm 0.10$\\
\hline
$D_s^+ \to K_S^0\pi^+$       & 673 & $+5.45 \pm 2.50 \pm 0.33$ & $\pm 0.30$\\
$D_s^+ \to K_S^0K^+$         & 673 & $+0.12 \pm 0.36 \pm 0.22$ & $\pm 0.10$\\
\hline
\end{tabular}
\caption{Projected Belle~II sensitivities, relative to existing Belle
measurements, for direct CP asymmetries in selected $D$ and $D_s$ 
decay modes.  When two uncertainties are given separately, they are
statistical and systematic.  Otherwise, the uncertainty is a combined
uncertainty.  Modes without a Belle~II sensitivity can be studied at 
Belle~II, but detailed sensitivity studies have not been conducted at
this time.}
\label{tab-Acp-BelleII}
\end{center}
\end{table}

Belle~II is well situated to make many of these measurements, and can
make uniquely useful contributions to modes with neutral particles
in the final state, complementary to those states accessible
at LHCb.  A list of selected Belle measurements and
prospects for improved measurements at Belle~II is given in
Table~\ref{tab-Acp-BelleII}.  As with the previously described projections,
the Belle~II values incorporate improvements due to statistics as well
as those systematic errors that are expected to improve with statistics.
As these further measurements become available, the necessity of new physics 
to explain direct CP asymmetries of $D$ meson decays may be revealed.

\section{Conclusions}

Upgrades at Belle~II and SuperKEKB are now underway, with physics operation 
planned to begin in 2016, and a dataset of $50~\mathrm{ab}^{-1}$ expected 
by 2022.  This large data sample will include
more than $6\times10^{10}$ charm events, allowing excellent sensitivity to 
many $D$ meson mixing and CP violation parameters.  In addition to the obvious
benefits afforded by a factor of 50 increase in statistics, a key element
of the Belle~II charm program is its large breadth.  Furthermore, many of the 
modes accessible at Belle~II, such as decays with neutral final state particles
or missing energy, are challenging or impossible in a hadron collider
environment.  Thus, the Belle~II program
is nicely complementary to those underway at the LHC.  Since mixing parameters 
and CP violation studies, both indirect and direct, require a 
large number of modes to make definitive conclusions about whether new 
physics is present, this complementarity of Belle~II measurements and others
will be a key element in advancing these searches for new physics.


\end{document}